\pdfoutput=1

\documentclass[aps,prd,twocolumn,nofootinbib,amsmath,amssymb,superscriptaddress,10pt]{revtex4}

\usepackage{amssymb,amsmath,graphicx,color,microtype,hyperref}

\usepackage{setspace}
\usepackage{leftidx}

\def \der{{\rm d}}

\def \nt{n_{\rm t}}
\def \at{\alpha_{\rm t}}

\def\apj{The Astrophysical Journal}

\def\aap{Astronomy and Astrophysics}

\def\jcap{Journal of Cosmology and Astroparticle Physics}
\def\prd{Physical Review D}

\def\apjs{The Astrophysical Journal Supplement Series}

\begin{document}

\title{How much primordial tensor mode is allowed?}
\author{Moumita Aich}
\email{aich@ukzn.ac.za} 
\affiliation{School of Mathematics, Statistics and Computer Science, University of KwaZulu-Natal, Durban, South Africa}
\author{Yin-Zhe Ma}
\email{ma@ukzn.ac.za} 
\affiliation{School of Chemistry and Physics, University of KwaZulu-Natal, Durban, South Africa}
\affiliation{NAOC-UKZN Computational Astrophysics Centre (NUCAC), University of KwaZulu-Natal, Durban, 4000, South Africa}
\author{Wei-Ming Dai}
\email{daiw@ukzn.ac.za} 
\affiliation{School of Chemistry and Physics, University of KwaZulu-Natal, Durban, South Africa}
\affiliation{NAOC-UKZN Computational Astrophysics Centre (NUCAC), University of KwaZulu-Natal, Durban, 4000, South Africa}
\author{Jun-Qing Xia}
\email{xiajq@bnu.edu.cn}
\affiliation{Department of Astronomy, Beijing Normal University, Beijing 100875, China}


\begin{abstract}
The presence of a significant amount of gravitational radiation in the early Universe affects the total energy density and hence the expansion rate in the early epoch. In this work, we develop a physical model to connect the parameter of relativistic degree of freedom $N_\mathrm{eff}$ with the amplitude and shape of primordial tensor power spectrum, and use the CMB temperature and polarization data from {\it Planck} and BICEP2/KECK Array, and the primordial deuterium measurement from damped Lyman-$\alpha$ (DLA) systems to constrain this model. We find that with this extra relation $\Delta N_{\rm eff}(r,n_{\rm t})$, the tensor-to-scalar ratio $r$ is constrained to be $r<0.07$ ($3\sigma$ C.L.) and the tilt of tensor power spectrum is $\nt=-0.01\pm 0.31$ ($1\sigma$ C.L.) for {\it Planck}+BICEP2+KECK+[D/H] data. This achieves a much tighter constraint on the tensor spectrum and provides a stringent test for cosmic inflation models. In addition, the current constraint on $N_{\rm eff}=3.122 \pm 0.171$ excludes the possibility of fourth neutrino species at more than $5\sigma$ C.L.
\end{abstract}

\maketitle

\section{Introduction}
\label{sec:intro}
The detection of large angular scale B-mode polarization of the Cosmic Microwave Background (CMB) is one of the next challenges in modern observational cosmology. The B-mode polarization signal arises only from tensor perturbations in the early Universe and is a direct signature of inflationary gravitational waves (GWs). The tensor perturbation power spectrum can be parametrized as
\begin{eqnarray}
P_{\rm t}(k) &= & A_{\rm t}(k_{0})\left(\frac{k}{k_{0}}
\right)^{\nt+(\at/2)\ln(k/k_{0})} \nonumber \\
&=& r A_{\rm s}(k_{0})  \left(\frac{k}{k_{0}}
\right)^{\nt+(\at/2)\ln(k/k_{0})} 
 \label{eq:Pt}
\end{eqnarray}
where $A_{\rm s}(k_{0})$ is the primordial scalar fluctuation amplitude, $r(k_{0})=A_{\rm t}(k_{0})/A_{\rm s}(k_{0})$ is the tensor-to-scalar ratio at a pivot scale $k_0$, $\nt$ is the spectral index of tensor power spectrum and $\at$ is the running of spectral index. 
%
For the single-field slow-roll inflation model the generic consistency relation $r=-8n_{\rm t}$ is satisfied. A power spectrum with a small negative tilt (red-tilt, $\nt < 0 $) is thus a characteristic prediction for single-field slow-roll inflation models~\cite{copeland}. Testing this prediction by using the data from CMB and Big-Bang nucleosynthesis (BBN) is an essential task to pin down the uncertainty of inflationary models. 

The consistency relation is however not satisfied for multi-field inflation and models which deviate from slow-roll. Alternative cosmological models, for example, string gas cosmology \cite{Brandenberger}, super-inflation models \cite{baldi} and many others not yet ruled out by observations, predict a blue tilt ($\nt > 0$) of the GW spectrum, i.e. more power at small scales. Therefore, observational constraints on the tilt of the tensor spectrum would be worth investigating since it has the distinguishing power in model space~\cite{Stewart, Brandenberger, Boyle, Lehners, Kuroyanagi}. It is thus appropriate to have a phenomenological approach by relaxing the consistency relation. Even though a direct detection of the inflationary GWs background is yet to be achieved, the current CMB measurement from {\it Planck}, Background Imaging of Cosmic Extragalactic Polarization (BICEP2) and KECK array data are already able to constrain it to a certain level~\cite{planck2015, planck2015_data, planck2018, bicep, bicep2018}. Apart from the CMB, other observational techniques also provide constraints on the stochastic GWs at different frequencies, through BBN~\cite{Allen, Maggiore}, pulsar timing~\cite{pta, pta-ska, meerburg, Boyle1}, and more directly through the Laser Interferometer Gravitational-Wave Observatory (LIGO) and Virgo interferometer GW detectors~\cite{ligo}.

\begin{figure}
\includegraphics[width=9cm]{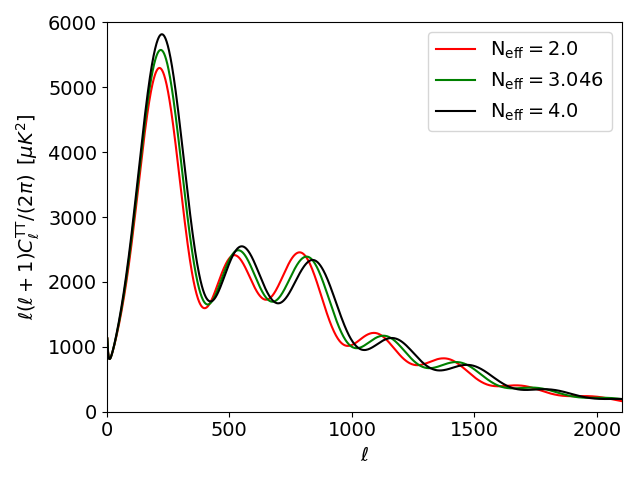}
\caption{The effect of $N_\mathrm{eff}$ (number of relativistic species) on the CMB temperature power spectrum, while fixing all other cosmological parameters as $\Omega_{\rm b}h^{2}=0.0224$, $\Omega_{\rm c}h^{2}=0.1201$, $\Omega_{\nu}h^{2}=0.6451 \times 10^{-3}$, $\Omega_{k}=0$, $H_{0}= 67.32\,{\rm km}\,{\rm s}^{-1}\,{\rm Mpc}^{-1}$, and $A_{\rm s}=2.1\times 10^{-9}$.}
\label{fig:cls}
\end{figure}

A blue-tilted tensor power spectrum would lead to additional small-scale relativistic degrees of freedom \cite{Kuroyanagi, meerburg, Stewart, Giovannini}, changing the energy density of the universe, which in turn would affect the expansion rate during that era. Relativistic neutrinos also contribute to the energy density of the universe, and the modification of the neutrino energy density can be parametrized by the effective number $N_\mathrm{eff}$ of neutrino species. The effect of the tensor blue tilt is thus degenerate with $N_\mathrm{eff}$. The Standard Model of particle physics predicts a $N_{\nu}=3.046$, so any extra value of the $N_{\rm eff}$ other than $N_{\nu}$ can be attributed either to an additional species of neutrino, or gravitational wave background. This radiation density has major ramifications on various early universe physical processes, leaving detectable imprints on the CMB at the epoch of last scattering. Fig.~\ref{fig:cls} shows the effect of the parameter $N_\mathrm{eff}$ on the CMB temperature anisotropies. The major physical effects are as follows
\begin{itemize}

\item Delaying matter-radiation equality~\cite{snowmass,Costanzi,simha,Dodelson,Komatsu09} -  As $N_{\rm eff}$ increases, the fractional density of radiation increases, therefore the matter to radiation equality occurs later. The amount of early ISW effect changes if the matter-radiation equality epoch changes. The earlier of matter-to-radiation epoch is, the more ISW effect that CMB photons receive~\cite{Komatsu09, Dodelson}. The effect can be measured through the ratio between the heights of the third and first acoustic peak of $C^{\rm TT}_{\ell}$, leading to the extraction of $z_{\rm eq}$ directly via CMB power spectrum~\cite{Komatsu09}. By using the relation of the present-day neutrino temperature and CMB temperature as $T_{\nu}=\left(4/11 \right)^{1/3}T_{\gamma}$, one can derive the equality epoch as~\cite{Dodelson,Komatsu09}
\begin{eqnarray}
1+z_{\rm eq}=\frac{\Omega_{\rm m}h^{2}}{\Omega_{\gamma}h^{2}}\frac{1}{1+0.2271\,N_{\rm eff}}, \label{eq:zeq}
\end{eqnarray}
where the radiation energy density $\Omega_{\gamma}h^{2}=2.47\times 10^{-5}$, and the present-day CMB temperature is $T_{\gamma}=2.725\,$K. As one can see from Eq.~(\ref{eq:zeq}), $\Omega_{\rm m}h^{2}$ and $N_{\rm eff}$ are linearly correlated with each other with the width of degeneracy dependent on the uncertainty of $z_{\rm eq}$. The anisotropic stress of relativistic degree of freedom can break the degeneracy, by imprinting distinct features on the CMB sky, independent of $\Omega_{\rm m}h^{2}$.


\item Adding anisotropic stress - Acoustic stress of the relativistic particle adds to the gravitational potential as an additional source of energy via Einstein's equation~\cite{Komatsu09}. In comparison, those relativistic particles which do not stream freely, but interact with matter frequently, do not have significantly anisotropic stress, because they isotropize themselves via interacting with matter. Therefore, the anisotropic stress of photons before decoupling time is very small. However, neutrino and graviton decoupled from hot plasma very early on, so the anisotropic stress is significant at the decoupling epoch. This effect is uncorrelated with $\Omega_{\rm m}h^{2}$, therefore, it can break the degeneracy.

\item Changing the sound horizon - In the standard cosmology model, free-streaming neutrinos travel supersonically through the photon-baryon plasma since their decoupling ($T\sim 1\,$MeV), so they gravitationally pull the wavefronts of the plasma oscillation slightly ahead of the time than the case when neutrinos are absent~\cite{Follin15, Baumann17, baumann, HV19, Kreisch19}. Therefore, the free-streaming neutrino changes the phase of the CMB acoustic oscillation by shifting the power spectra towards larger angular scales (smaller $\ell$), while also suppressing the damping tail. Similar to neutrinos, any relativistic degree of freedom also has a similar effect of altering the scale of the sound horizon, therefore cause the distinctive shift in the CMB power spectra \cite{Follin15,Baumann17,baumann,HV19,Kreisch19}.

\end{itemize}

In Fig.~\ref{fig:cls}, we plot the comparison of CMB temperature power spectrum $C^{\rm TT}_{\ell}$ by varying the values of $N_{\rm eff}$ while fixing other cosmological parameters. One can clearly see the distinctive change of the power spectrum due to the combination of all three effects above. To see each effect, one needs to fix the sound horizon, or fix the $z_{\rm eq}$ and vary $N_{\rm eff}$. We refer to the interested readers to Fig.~1 in Ref.~\cite{HV19}, Fig.~1 in Ref.~\cite{Kreisch19}, Fig.~2 in Ref.~\cite{Baumann17}, and Fig.~1 in Ref.~\cite{Follin15}.

High precision CMB observations such as space-based {\it Wilkinson Microwave Anisotropy Probe} (\textit{WMAP})~\cite{Hinshaw13} and \textit{Planck} satellite~\cite{Planck18-legacy,planck2018}, ground-based Atacama Cosmology Telescope (ACT) \cite{act1,act2}, South Pole Telescope (SPT) \cite{spt1,spt2}, BICEP2-KECK Array \cite{bicep, bicep2018} and balloon-based (SPIDER) \cite{spider1, spider2, spider3} experiments have the potentiality to give rigorous constraints on the neutrino background. Therefore, these experiments should also be able to place strong constraints on extra relativistic degree of freedom caused by blue tilted tensor power spectrum. In this paper we firstly explore to constrain the effective relativistic species using current CMB data from \textit{Planck}, and the BICEP2/KECK array, characterising this case as the standard ``CMB with $\Delta N_\mathrm{eff}$ relation'' case.


Besides the effect in the CMB, relative abundances of primordial light elements such as hydrogen (H), deuterium (D), helium-3 ($\leftidx{^3}{\mathrm{He}}$), helium-4 ($\leftidx{^4}{\mathrm{He}}$) and small amounts of lithium-7 ($\leftidx{^7}{\mathrm{Li}}$) created during the BBN, are also strongly affected by the GW background. Significant gravitational radiation during primordial nucleosynthesis affects the total energy density of the universe, resulting in altering the expansion rate of the universe. Thus the relative abundances of the light elements thus would vary from the predictions from standard BBN if the GW background is modified. This is an indirect constraint on the energy density of the GW background~\cite{Stewart}.


Constraints of $N_\mathrm{eff}$ from CMB measurements are mostly derived from measurements of the damping tail \cite{planck2015,spt_damping, nu_damping}. An increase in the radiation density of the early universe reduces the mean free path of fluctuations in the photon baryon fluid and increases the damping of small-scale fluctuations. Changes to the helium and deuterium fraction ($Y_\mathrm{p}$ and D/H) induces a variation in the free electron fraction which in turn alters the mean free path of the photons and affects the damping tail \cite{NollettHolder2011}. We use the D/H measurements from \cite{cooke2016, cooke2018} complemented with \textit{Planck} and BICEP2/KECK likelihoods to study the constraints on the $N_\mathrm{eff}$ and consequently the effect on the $r-n_{\rm t}$ joint distribution, characterising this case as the standard ``CMB + D/H with $\Delta N_\mathrm{eff}$ relation'' case. 

The paper is organized in the following way. Section~\ref{dof} discusses the relation between the primordial GW energy density and the effective degree of freedom of relativistic species. Then we discuss how does this $\Delta N_{\rm eff}$ impacts on helium production. Section~\ref{data} introduces the datasets we use for our analysis, i.e. CMB data from the \textit{Planck} satellite and the BICEP2/KECK array 2018 release. We also use deuterium abundance data from DLAs which serve as an independent measurement of the $N_\mathrm{eff}$. Section~\ref{results} presents the results of our Markov-Chain Monte-Carlo runs and their implication. The conclusion and future goals are presented in Section~\ref{conclusions}. Throughout the paper, we adopt a spatially-flat $\Lambda$CDM cosmology model with adiabatic initial conditions.


\section{Relativistic degrees of freedom}
\label{dof}
Stochastic GW background searches venture to measure the fractional energy density of GWs as a function of frequency. We define the logarithmic GW contribution to the critical density as~\cite{Maggiore00, Smith06}
\begin{eqnarray}
\Omega_{\rm GW}(k) &\equiv& \frac{1}{\rho_{\rm c}}\frac{\der \rho_{\rm GW}}{\der \ln k}, 
\end{eqnarray}
where $\rho_\mathrm{GW}$ is the frequency (wave-number $k$) dependent effective energy density, and $\rho_{\rm c} = 3c^2 H^2_0/8 \pi \mathrm{G}$ is the critical density of the Universe at present, with $H_{0} \equiv 100h\,{\rm km}\,{\rm s}^{-1}\,{\rm Mpc}^{-1}$ is the current Hubble parameter. The GW energy density can be related to $h_k$, which is the Fourier transform of the metric perturbation as 
\begin{eqnarray}
\Omega_{\rm GW}h^{2} &=& \frac{c^{2}k^{2}h^{2}}{6H^{2}_{0}}\langle |h^{2}_{k}| \rangle \equiv A_{\rm GW}P_{\rm t}(k),
\label{eq:omega-gw1}
\end{eqnarray}
where $A_{\rm GW}=2.74 \times 10^{-6}g^{-1/3}_{100}$, where $g_{100} \equiv g_{\ast}(T_{\rm k})/100$ is the degree of freedom at the time when GWs stretched out of the Hubble radius. If counting only standard-model particles, $g_{\ast}(T_{k})= 106.75$ so $g_{100}=1.06$~\cite{Kolb-Turner}.

In the early Universe before BBN, graviton behaves like relativistic particles whose density $\sim a^{-4}$. Thus if there were too many gravitons before BBN, it would enhance the total energy density of the Universe substantially, therefore, making the Universe expand too fast. Firstly, we need to calculate what was the GW density back to the time of BBN. 

\subsection{Energy densities}
The energy densities for neutrinos, gravitons and photons are given as 
\begin{eqnarray}
\rho_{\nu} &=&
\frac{7}{8}\frac{\pi^{2}}{15}N_{\nu}T^{4}_{\nu} \nonumber \\
\rho_{\rm G} &=& \frac{7}{8}\frac{\pi^{2}}{15} \Delta N_{\rm eff} T^{4}_{\nu} \nonumber \\
\rho_{\gamma} &=& \frac{\pi^{2}}{15} T^{4}_{\gamma},
\label{eq:density}
\end{eqnarray}
in which we assume the graviton spin is 2. The neutrino temperature is related to the CMB temperature as $T_{\nu}=(4/11)^{1/3}T_{\gamma}$, where we take $T_\gamma=2.726$ K. Back to BBN time
\begin{eqnarray}
\rho_{\rm c}(t_{\rm n}) &=& \rho_{\gamma}(t_{\rm
n})+\rho_{\nu}(t_{\rm n}), \\
\frac{\rho_{\rm c}(t_{\rm n})}{\rho_{\gamma}(t_{\rm
n})}&=&\frac{N_{\nu}\frac{7}{8}\left(\frac{4}{11}\right)^{4/3}+1}{1}
\simeq 1.692.
\end{eqnarray}
We integrate Eq.~(\ref{eq:omega-gw1}) over all possible scales and use Eq.~(\ref{eq:density}) to calculate the increment of the effective number of relativistic species back to BBN time
\begin{eqnarray}	
\Delta N_{\rm eff} &=& \frac{8}{7} \left(\frac{11}{4}\right)^{4/3} \frac{A_{\rm GW}}{h^{2}}\frac{\rho_{\rm c}(t_{\rm n})}{\rho_{\gamma}(t_{\rm n})} {\displaystyle \int\limits^{\ln k_{\rm max}}_{\ln k_{\rm min}}} \!\!\! \der \ln k ~ P_{\rm t}(k) \nonumber \\
&=& 1.781 \times 10^{5} A_{\rm GW} {\displaystyle \int\limits^{\ln k_{\rm max}}_{\ln k_{\rm min}}} \!\!\! \der \ln k ~ P_{\rm t}(k) \,.
\label{eq:neff}
\end{eqnarray}

\subsection{The integral}
To evaluate the integral in Eq.~(\ref{eq:neff}), we need to figure out the upper and lower limits of the wave-number $k$. We set $k_{\rm min}$ as the particle horizon of the Universe at 
BBN time, so $k_{\rm min} \simeq k_{H_{\rm BBN}}$. 
$k_{\rm max}$ corresponds to the minimal scales of the perturbation, which entered into the Hubble radius right after inflation, so assume $k_{\rm max}=e^{60}k_{H_{0}}$. 

\begin{eqnarray}
k_{H_0} &=&\frac{2\pi}{\left(\frac{c}{H_0}\right)}=\frac{2\pi
H_{0}}{c} = 0.0015\, {\rm Mpc}^{-1},
\end{eqnarray}
if we assume $H_0=70 \,{\rm km}\,{\rm s}^{-1}\, {\rm Mpc}^{-1}$.

\begin{eqnarray}
k_{\rm min}&=& \frac{2\pi}{\left(\frac{c}{H_{\rm BBN}a_{\rm
BBN}}\right)}=\frac{2\pi H_{\rm BBN}a_{\rm BBN}}{c}. \\ \nonumber
\end{eqnarray}
\begin{eqnarray}
 H^{2}_{\rm
BBN} & = & \frac{8 \pi G}{3}(\rho_{\gamma}+\rho_{\nu}) \nonumber
\\ &=&\frac{8\pi G}{3}\rho_{\gamma}(1+0.2271N_{\rm eff})
 \nonumber \\
& \simeq & \frac{8\pi G}{3} \times 1.69 \times \rho_{\gamma}
\end{eqnarray}
where for the second line we use Eq.~(\ref{eq:density}). Since
\begin{eqnarray}
\frac{\rho_{\gamma}}{\rho_{\rm cr}}=\frac{2.47\times
10^{-5}}{h^{2}a^{4}} ,
\end{eqnarray}
we have
\begin{eqnarray}
H_{\rm BBN}=H_{0} \times \left[\frac{6.46 \times 10^{-3}}{h a^{2}}
\right].
\end{eqnarray}
Therefore, we have
\begin{eqnarray}
k_{\rm min} &=& 2\pi \left(\frac{H_{\rm BBN}a_{\rm BBNc}}{c}
\right)  \nonumber \\ &= & \left(2\pi \frac{H_{0}}{c} \right)\left(
\frac{6.46 \times 10^{-3}}{h a_{\rm BBN}} \right),
\end{eqnarray}
where $a_{\rm BBN}=T_{0}/T_{\rm BBN} \simeq 2.275 {\rm K}/ 1{\rm
MeV}=1.96 \times 10^{-10}$ and $h=0.7$. Thus
\begin{eqnarray}
k_{\rm min}=4.7 \times 10^{7} \left(2\pi \frac{H_{0}}{c}
\right) = 6.9 \times 10^4 ~\mathrm{M pc^{-1}}~.
\end{eqnarray}

Now let us focus on calculating the integral
\begin{eqnarray}
I_{1}(\nt,\at)=\int \der \ln k \left(\frac{k}{k_{0}} \right)^{\nt
+(1/2) \at \ln (k/k_{0})},
\end{eqnarray}
which is found to be
\begin{eqnarray}
 I_{1}(\nt,\at) &=& e^{-(n_{\rm t}/\alpha_{\rm t})^{2}/2}\left(\frac{\pi}{-2 \at} \right)^{1/2}  \nonumber \\ &\times&
 \left[{\rm
Erf}\left(\sqrt{-\frac{\at}{2}}\left(\ln \left(\frac{k_{\rm
max}}{k_{0}} \right) +\frac{\nt}{\at} \right) \right) \right. 
\nonumber \\ &-& \!\! \! \left. {\rm Erf}\left(\sqrt{-\frac{\at}{2}}\left(\ln
\left(\frac{k_{\rm min}}{k_{0}} \right) +\frac{\nt}{\at} \right)
\right) \right] \! , \label{eq:I1}
\end{eqnarray}
where ``Erf'' represents for Error Function.


\begin{figure}[!]
\includegraphics[width=3.3in]{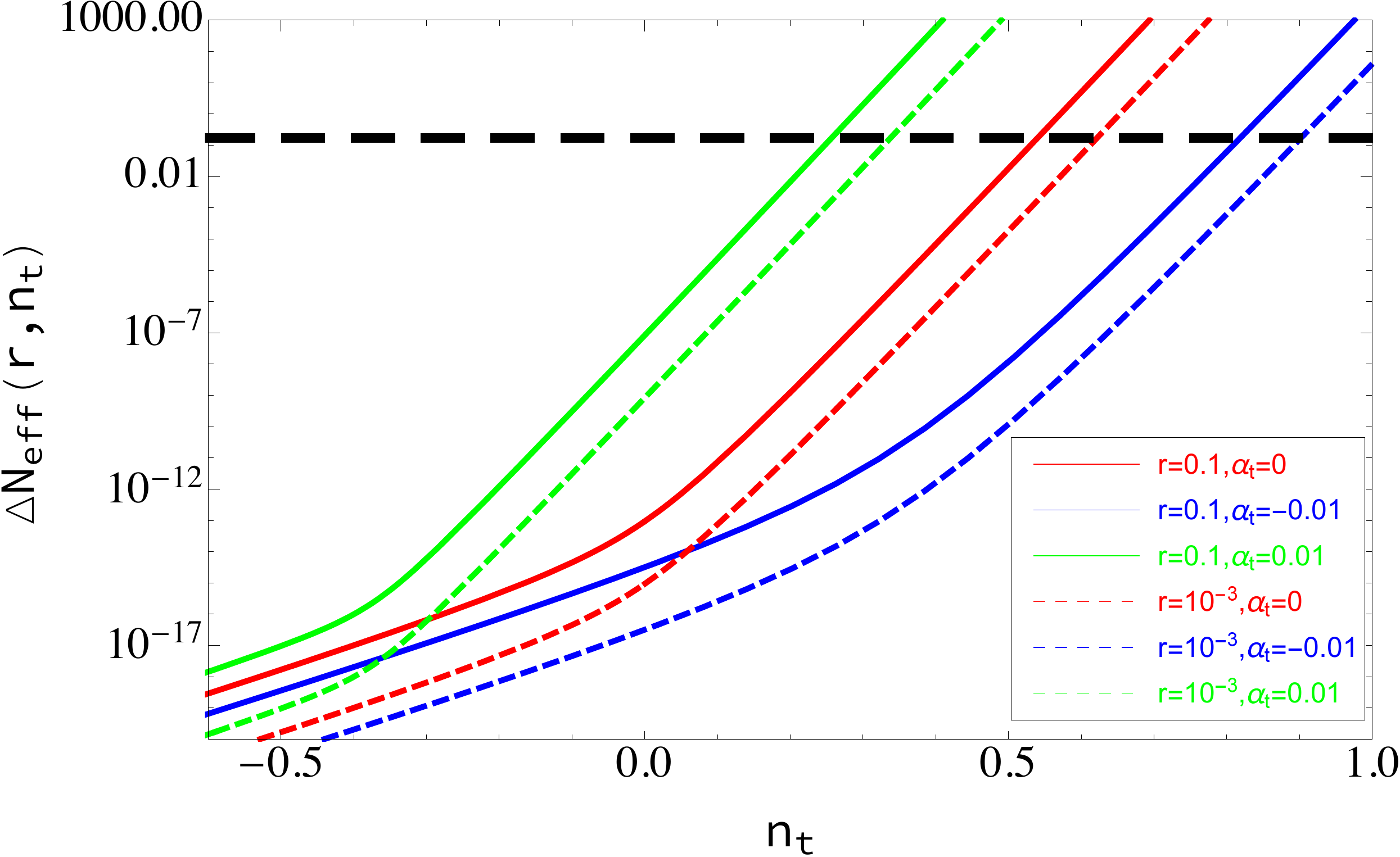}
\caption{The increment of effective number of relativistic species
given the value of $\nt$. The three curves show the assumed value
of $\at$. The horizontal dashed line shows the current $1\sigma$ limit of $\Delta N_{\rm eff}$~\cite{planck2018}.} \label{fig:nt}
\end{figure}

We consider $\at=0$, given that the current CMB data does not point to any strong evidence of the running of the tensor tilt. Combining the above equations, we find
\begin{eqnarray}
\Delta N_{\rm eff}(r,\nt)=1.781 \times 10^{5}A_{\rm GW}A_{\rm
s}(k_{0})r I_{1}(\nt)~, \label{eq:delta-Neff}
\end{eqnarray}
where
\begin{eqnarray}
I_{1}(\nt) &=& \frac{1}{\nt} \left[\left(\frac{k_{\rm max}}{k_0} \right)^{\nt}- \left(\frac{k_{\rm min}}{k_0} \right)^{\nt} \right] ,\, {\rm if}\,\,{\nt \neq 0} \nonumber \\
&=& \ln\left(\frac{k_{\rm max}}{k_{\rm min}} \right),\, {\rm if}\,\,{\nt = 0}.
\end{eqnarray} 
Here
\begin{eqnarray}
k_{\rm max} &=& {\rm e}^{60}k_{H_{0}} \simeq 1.67 \times 10^{23}\,{\rm Mpc}^{-1}, \nonumber \\
k_{\rm min} &=& 6.9 \times 10^{4} \,{\rm Mpc}^{-1},
\end{eqnarray}
and we take the pivot scale as $k_{0}=0.01\,{\rm Mpc}^{-1}$ to be consistent with {\it Planck}~\cite{planck2018}. Figure~\ref{fig:nt} illustrates the relation of $\Delta N_{\rm eff}$ as a function of $\nt$ for different values of $r$ and $\at$. We plot the current $95\%$ confidence level (C.L.) upper limit of $r$ as $r<0.1$ from the constraints of {\it Planck} TT+TE+EE+lowE+lensing~\cite{planck2018}, and the cosmic variance limit of $r$ as $r=10^{-3}$~\cite{Ma10}. The horizontal black dashed line shows the $\Delta N_{\rm eff}<0.17$ as the $68\%$ C.L. upper limit~\cite{planck2018}. One can see that, even if the value of $r$ is small, the blue tilted $\nt$ can lead to large increment of $\Delta N_{\rm eff}$, resulted in the observable effect in the CMB and light element abundance.  

\subsection{Helium abundance}

\begin{figure}[!t]
\includegraphics[width=3.2in]{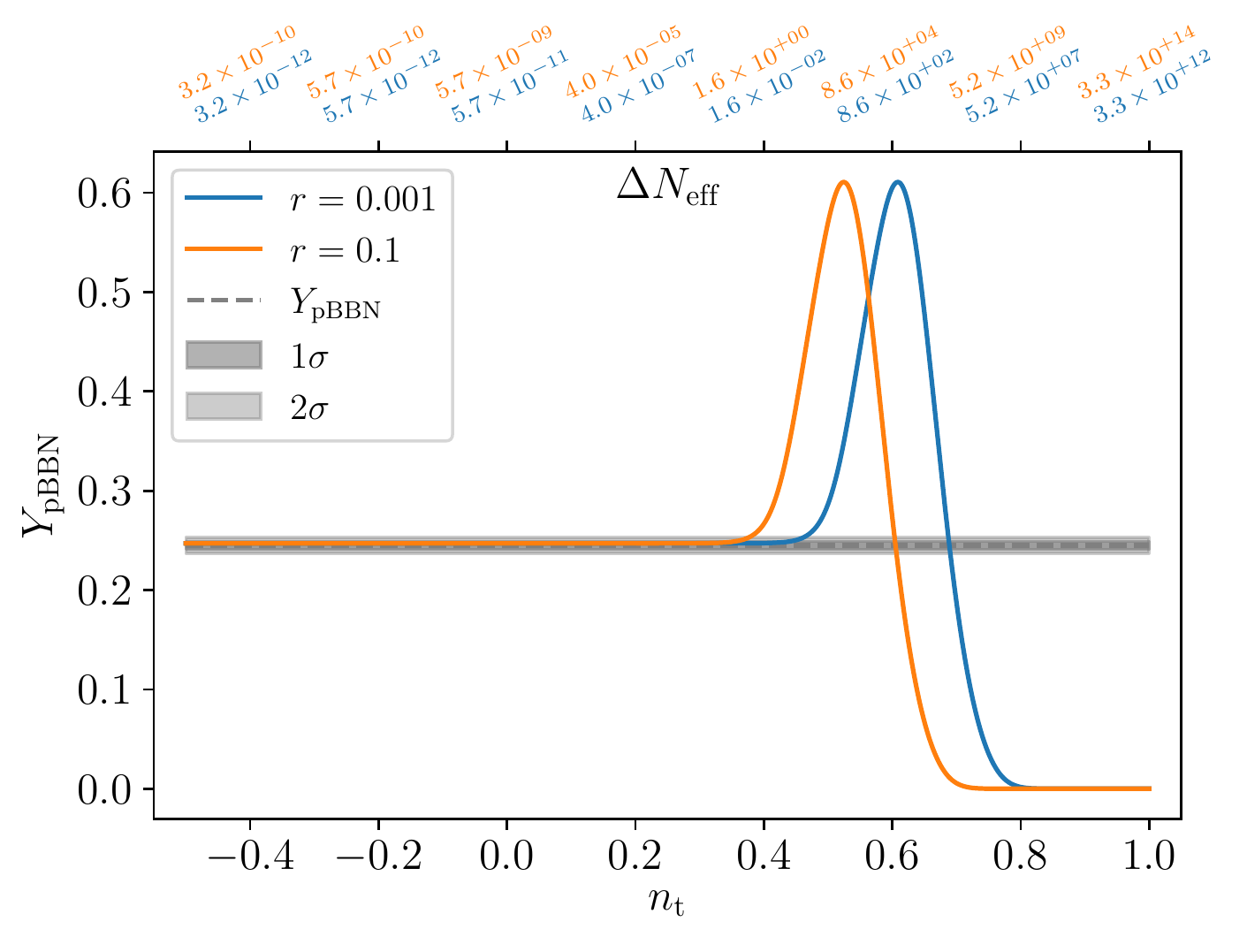}
\caption{The abundance of helium ($Y_{\rm p}$) as a function of $\nt$. The blue and orange lines are for $r=10^{-3}$ and $r=0.1$ respectively. The horizontal grey band is the currently measurement $Y_{\rm p}$ (Eq.~(\ref{eq:helium-yp})) from~\cite{aver1}.} \label{fig:nt_He}
\end{figure}

Given a set of cosmological parameters, the primordial abundance of light elements is fully computable from the standard model of particle physics~\cite{Wagoner}. The precise determination of cosmological parameters from {\it Planck} satellite leads to accurate prediction of the light element abundance, such as $\leftidx{^4}{\mathrm{He}}$ from low metallicity H$_\mathrm{II}$ regions in low-redshift star-forming galaxies~\cite{aver1,aver}, primordial abundance of deuterium (D/H) using quasar absorption lines like the DLAs \cite{cooke2013, cooke2016, cooke2018}, and $\leftidx{^7}{\mathrm{Li}}$/H ratio in metal-poor stars in the Milky Way halo~\cite{Sbordone10}. The standard BBN populations of relativistic particles, including photons, electrons, positrons, and three species of neutrinos mix as a hot plasma with the same temperature. At a given temperature, the resulting cosmic expansion rate is $2.3$ times that of photon alone. The weak freeze-out starts at this time, settling down the neutron-to-proton ratio which eventually determines the helium abundance $Y_{\rm p}$. Additional relativistic degree of freedom can enhance the expansion rate by a factor of $8\%$\footnote{The same value of enhanced expansion rate due to an additional neutrino species quoted in sec.~2.B. in Ref.~\cite{NollettHolder2011} is $40.3\%$, which we believe to be an error.}, which forces the neutrino freeze-out to occur at a higher temperature. This, in turn, implies more neutrons, triggering more $\leftidx{^4}{\mathrm{He}}$. 

By modifying the publicly available {\tt PArthENoPE} 2.0 code\footnote{\url{http://parthenope.na.infn.it}}, we implement the $\Delta N_{\rm eff}(r,n_{\rm t})$ relation into the code and output the helium abundance as a function of $\nt$ by fixing the $r$ value. In Fig.~\ref{fig:nt_He}, we plot the helium abundance as a function of $\nt$ for the cases of $r=10^{-3}$ and $r=0.1$. The $\Delta N_{\rm eff}$ values are marked in the upper boundary of the horizontal axis. One can see that as $\nt$ goes from negative value to slightly positive value, the $\Delta N_{\rm eff}$ increases dramatically, leading to a higher temperature of neutrino freeze-out. This will lead to more helium production. However, there is a downward branch of $Y_{\rm p}$ when $\nt \gtrsim 0.5$. This is because, if the $\nt$ becomes very positive, the $N_{\rm eff}$ value becomes exponentially large, then the Hubble expansion at the early Universe becomes too fast to allow nucleons to interact and form helium. The Universe would then become cooling down too soon for the nucleus to synthesize into helium. Therefore, for $\nt \gtrsim 0.5$, i.e. $\Delta N_{\rm eff} \gtrsim 10^{3}$, the helium production becomes much lower than standard model.

The current measurement on helium abundance from observations of the helium and hydrogen emission line from metal-poor extragalactic H$_\mathrm{II}$ region, combined with estimated metallicity, give the primordial helium abundance as~\cite{aver1}
\begin{eqnarray}
Y_{\rm p}= 0.2449 \pm 0.0040 \,, \label{eq:helium-yp}
\end{eqnarray}
which is shown as the horizontal grey band in Fig.~\ref{fig:nt_He}. Large systematic uncertainties and degeneracies among the input parameters needed to model emission line fluxes limit the measurements predominantly. Along with this large error-bar predicament, this also results in a $2\sigma$ deviation from the standard $\Lambda$CDM prediction \cite{olive, aver, aver1}. More recently, the determination of $Y_{\rm p}$ from the measurement of the absorption feature of the intergalactic gas cloud against the light of a background quasar was made~\cite{cooke2018}, though the measurement error is still quite large. Due to these reasons, in the next section, we will only use the measurement of deuterium abundance to constrain $\Delta N_{\rm eff}$ relation.

%

%
%


\section{Data}
\label{data}
\subsection{Deuterium abundance}
\label{deut}

The deuterium abundance is also closely related to the number of relativistic species that existed during BBN. The abundance of deuterium is determined by the $d+d$ and $d+p$ reactions towards the end of BBN when the photon temperature drops below the rest mass of the electron. Therefore, essentially there is no electron and positron at this time, and they have been annihilated to heat the photon. The expansion rate at this time is $\sim 1.7$ times of the photon alone, and an additional $N_{\rm eff}$ will cause the speed up of cosmic expansion. This will lead to less time for deuterium burning and therefore end up in higher D/H~\cite{NollettHolder2011}.

The deuterium abundance is now more precisely measured by a significant factor compared to measurements of $Y_\mathrm{p}$, through the analysis of the most metal-poor damped Ly$\alpha$ (DLA) systems, which also displays the Lyman series absorption lines of neutral deuterium~\cite{cooke2013, cooke2016, cooke2018}. The primordial abundance of deuterium, on the other hand, has a monotonic response to $\Omega_{\mathrm{b},0} h^2$, and accurate measurements of the primordial D/H ratio complemented by measure of $\Omega_{\mathrm{b},0} h^2$ from the CMB, can provide a much more sensitive constraint upon allowed values of $N_\mathrm{eff}$ \cite{NollettHolder2011, cooke2013}. 


We use the D/H measurements from \cite{cooke2018} complemented with \textit{Planck} and BICEP2/KECK likelihoods to study the constraints on the $N_\mathrm{eff}$ and consequently the effect on the $r-n_{\rm t}$ joint distribution. The most recent measurement of cosmic deuterium abundance \cite{cooke2018} is derived from six damped Lyman alpha system and is given as
\begin{eqnarray}
10^{5}({\rm D}/{\rm H})_{\rm P}=2.527 \pm 0.030. \label{eq:DH}
\label{eq:deut}
\end{eqnarray}
We use the above measurements from \cite{cooke2018} as a primordial element abundance dataset and likelihood in CosmoMC. We also use the updated theory table with reduced errors from \cite{Marcucci_2016} using the publicly available \texttt{PArthENoPe} 2.0 code \cite{Pisanti}, for computing the abundances of light elements produced during BBN as a function of baryon density and number of radiation degrees of freedom. 

\begin{figure}[!]
\centerline{
\includegraphics[width=3.3in]{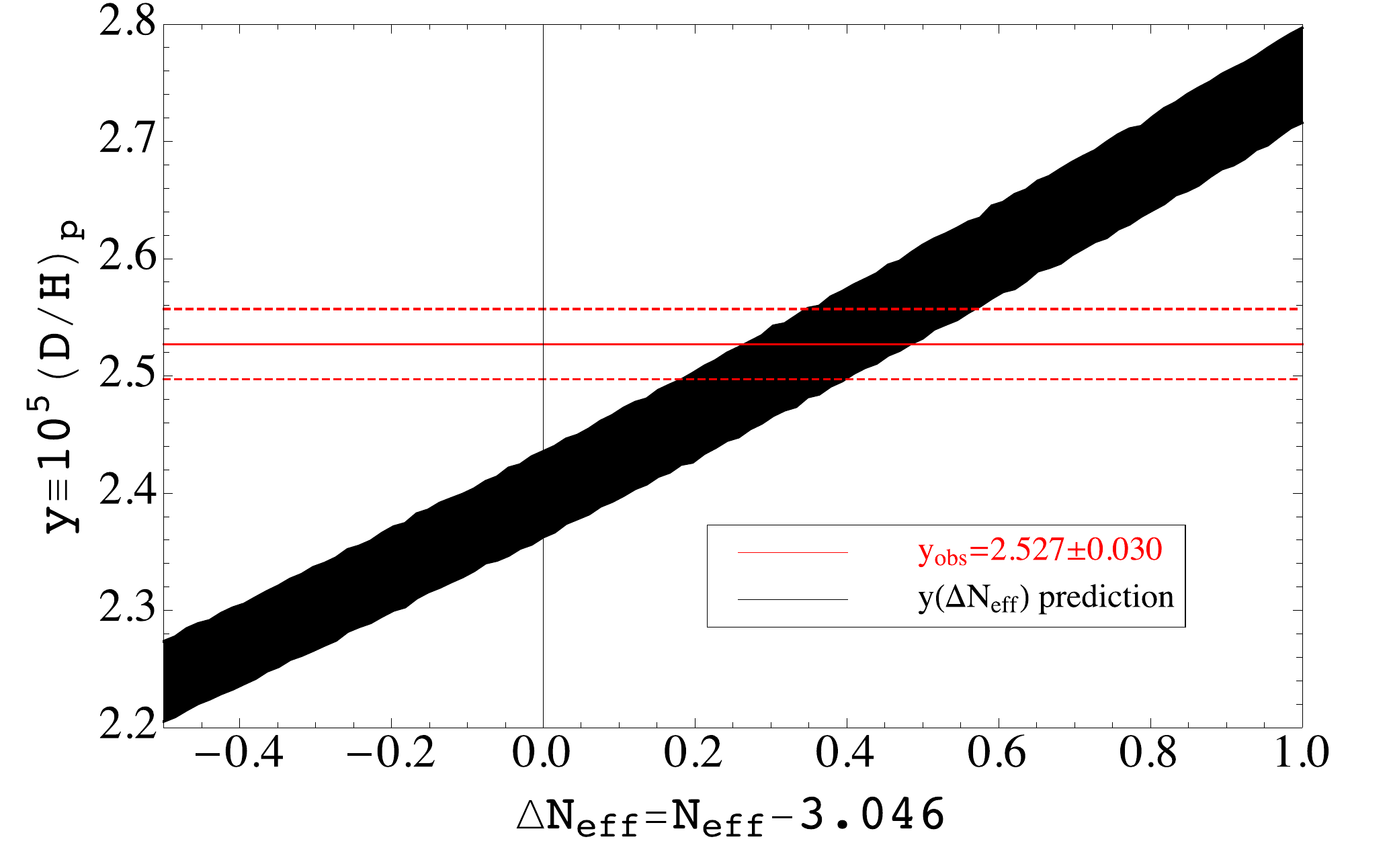}
}
\caption{The relationship between deuterium prediction and $\Delta N_\mathrm{eff}$.} 
\label{fig:yNeff}
\end{figure}

The relation between deuterium abundance and $\Delta N_{\rm eff}$ is (eqs.~(8)--(10) in~\cite{cooke2016})
\begin{eqnarray}
\eta_{\rm D}=6\left[\frac{10^{5}({\rm D}/{\rm H})_{\rm P}}{2.47} \right]^{-1/1.68}, \label{eq:etaD}
\end{eqnarray}
and
\begin{eqnarray}
\Delta N_{\rm eff}=\frac{43}{7}\left[\left(1+\frac{\eta_{10}-\eta_{\rm D}}{1.08(1.1\eta_{10}-1)} \right)^{2}-1 \right], \label{eq:Delta-neff2}
\end{eqnarray}
where
\begin{eqnarray}
\eta_{10} = (273.78 \pm 0.18) \times \Omega_{\rm b}h^{2}. \label{eq:eta10}
\end{eqnarray}
In Fig.~\ref{fig:yNeff}, we plot the relation between $\Delta N_{\rm eff}$ and the deuterium abundance $y\equiv 10^{5}({\rm D}/{\rm H})_{\rm p}$ by using Eqs.~(\ref{eq:etaD})--(\ref{eq:eta10}). We allow the $\Omega_{\rm b}h^{2}$ ($\Omega_{\rm b}h^{2}=0.02236 \pm 0.00015$ according {\it Planck} TT+TE+EE~\cite{planck2018}) and the front factor in Eq.~(\ref{eq:eta10}) to vary within $1\sigma$ C.L., shown as the width of the black band. One can see that the larger the $\Delta N_{\rm eff}$ is, the larger the prediction of the deuterium prediction. The reason is as follows. The deuterium abundance is determined by the $d(d,n)\leftidx{^3}{\mathrm{He}}$ and $d(p,\gamma)\leftidx{^3}{\mathrm{He}}$ processes that burn the deuterium, at the end of BBN. At this time, the photon temperature is around $T\sim 0.02\,$MeV, which is well below the electron rest mass. Therefore, electron-positron annihilated and increased the photon temperature, leading to the gap between neutrino temperature and photon temperature $T_{\nu}=(4/11)^{1/3}T_{\gamma}$. The expansion rate at this time in standard BBN is $1.3$ times that of photons alone, and an additional neutrino species at the same temperature as the others cause a $7\%$ speed-up. Faster expansion rate means the less time to burn deuterium, leading to higher value of D/H~\footnote{In sec.~2.B. in Ref.~\cite{NollettHolder2011} it is written that at the end of BBN, the expansion rate in standard BBN model is $1.7$ times that of photons alone, and an additional neutrino species causes $36.7\%$ speed-up, which we believe the numbers are wrong.}. The horizontal red lines are the $\pm 1\sigma $ measurement as Eq.~(\ref{eq:DH}). From Fig.~\ref{fig:yNeff}, one can see that $\Delta N_\mathrm{eff} \sim 0.3$ is preferred by the comparison between theory and the measurement. 
The D/H dataset described above is incorporated in CosmoMC through a supplementary likelihood and dataset with mean and error given as in Eq. (\ref{eq:deut}) and a theory table with D/H abundances as a function of $\Omega_{\rm b} h^2$ as given in PArthENoPE$\_$880.2$\_$marcucci.dat dataset in CosmoMC. We use this D/H dataset along with CMB datasets from {\it Planck} and BICEP2/KECK. 

\subsection{CMB data from {\it Planck} and BICEP2/KECK}
\label{data:cmb}
We use the standard cosmological Markov chain Monte Carlo (MCMC) package CosmoMC \cite{cosmomc} along with the \textit{Planck} 2015 likelihood \cite{planck2015_data} for our analysis. We use \textit{Planck} high-$\ell$, Plik TTTEEE, nuisance-marginalized likelihood in the range $\ell$=30-2508 for TT and $\ell$=30-1996 for TE and EE, and the low-$\ell$ TEB (TT, EE, BB and TE) likelihood in the range $\ell$=2-29. 

The BICEP2/KECK array (BK) currently operating at 95, 150 and 220 GHz, has the tightest upper limits to the tensor-to-scalar ratio $r$. For our initial analysis we were using BK 2015 data \cite{bicep} 
but with the release of BK 2018 dataset \cite{bicep2018} 
we completed our analysis using the latest dataset. We elaborate in Section~\ref{results} the potential tension between the BK 2015 and BK 2018 data in constraining the tensor parameters $r$ and $n_{\rm t}$, also illustrated in Fig.~\ref{fig:bkpver}. For the BK 2018 data, the BB band-powers are split into nine multipole bins from $\ell$=37-332 and contains a total of 12 auto and 66 cross spectra between BK 2018 maps at 95, 150, and 220 GHz, {\it WMAP} maps at 23 (K-band) and 33 GHz (Ka-band), and {\it Planck} maps at 30, 44, 70, 100, 143, 217, and 353 GHz \cite{bicep, bicep2018}. We use the above joint analysis BK 2018  and \textit{Planck} 2015 (henceforth BKP) data and likelihood provided with CosmoMC.


We allow the six standard cosmological parameters ($\Omega_{\rm b}h^{2}$, $\Omega_{\rm c}h^{2}$, $100\theta_{\ast}$, $\tau$, $\ln(10^{10}A_{\rm s})$, $n_{\rm s}$) to vary in our likelihood chain. We also release the tensor-to-scalar ratio $r$, the tilt of the tensor power spectrum $n_{\rm t}$ and effective number of relativistic species $N_{\rm eff}$ to vary in the likelihood, so essentially we have $9$ free parameters in total.

The default results are with only the {\it Planck} and the BKP CMB dataset mentioned above. We also combined the D/H data from Ly$\alpha$ forest to tighten up constraint on $N_{\rm eff}$. We modified CAMB/CosmoMC to introduce the effect of additional relativistic species $\Delta N_\mathrm{eff}$ to the GW background as discussed in Eq.~(\ref{eq:delta-Neff}), and then switch on-and-off this relation to test the additional constraints on the tensor power spectrum. Therefore, we have four different data sets, {\it Planck}, {\it Planck}+[D/H], BKP and BKP+[D/H]; and two models to fit, one is the $9$ cosmological parameter model without $\Delta N_{\rm eff}$ relation (Eq.~(\ref{eq:delta-Neff})) and the other is with this relation.

\section{Results of Constraints}
\label{results}

\begin{figure}[!h]
\centerline{
\includegraphics[width=2.8in]{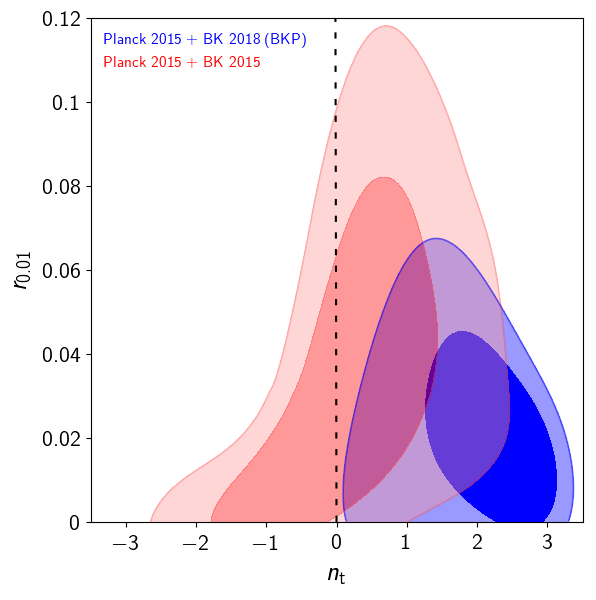}
}
\caption{The $2\sigma$ contour plot of the tensor to scalar ratio $r_{0.01}$ at $k=0.01\,{\rm Mpc}^{-1}$ and the tensor tilt $n_{\rm t}$ for {\it Planck} + BICEP2+KECK datasets, where the red and blue contours represent BICEP2/Keck 2015 and BICEP2/Keck 2018 dataset respectively.} 
\label{fig:bkpver}
\end{figure}

\begin{figure*}[!]
\centerline{
\includegraphics[width=2.15in]{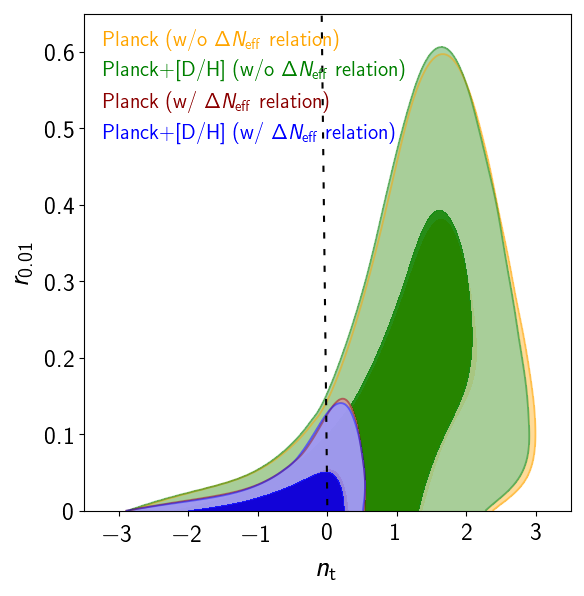}
\includegraphics[width=2.23in]{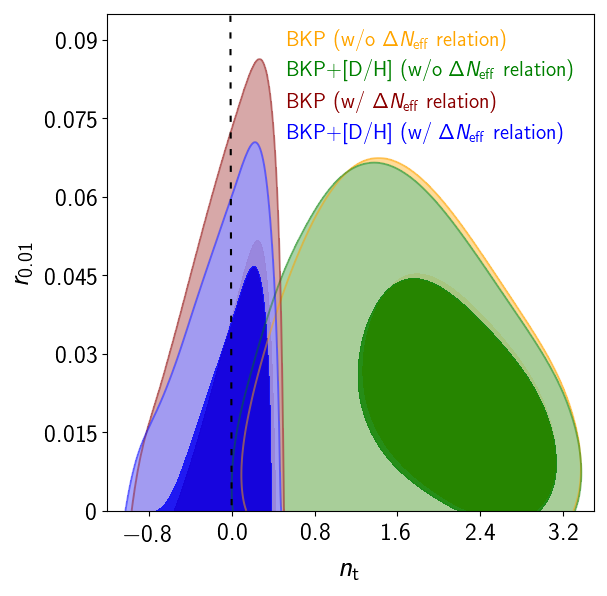}
\includegraphics[width=2.22in]{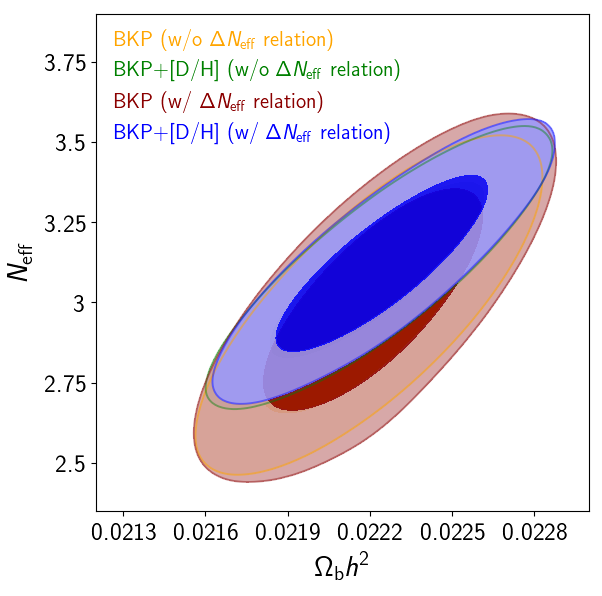}
}
\caption{$2\sigma$ contour plots for the various datasets and cases. The \textit{left} and the \textit{middle} plot shows the joint constraint of the tensor to scalar ratio $r_{0.01}$ and the tensor tilt $n_{\rm t}$ for {\it Planck} dataset and BICEP2+KECK+{\it Planck} (BKP) dataset respectively. The \textit{right} panel shows the joint constraint for $N_{\rm eff}$ and $\Omega_{\rm b} h^2$ for BKP dataset. (The results for {\it Planck} only dataset is not shown for redundancy.) For all plots, the four contours are for the joint constraint for the default case with CMB data only (yellow), CMB+[D/H] data (green), CMB data with $\Delta N_\mathrm{eff}$ constraint (red) and CMB+[D/H] data with $\Delta N_\mathrm{eff}$ constraint (blue).} 
\label{fig:2dfig}
\end{figure*}

\begin{figure*}[!]
\includegraphics[width=2.2in]{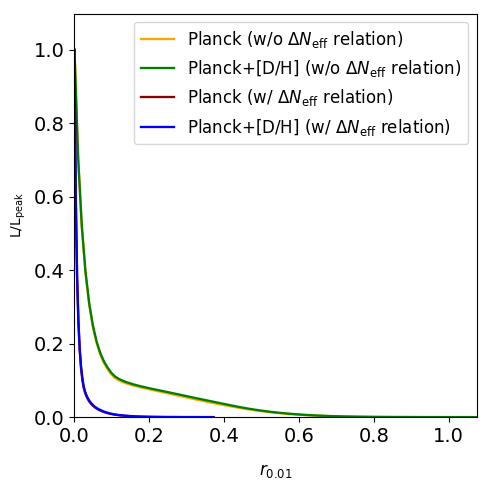}
\includegraphics[width=2.2in]{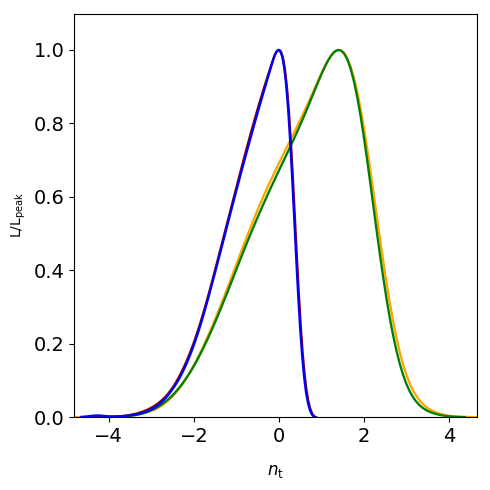}
\includegraphics[width=2.2in]{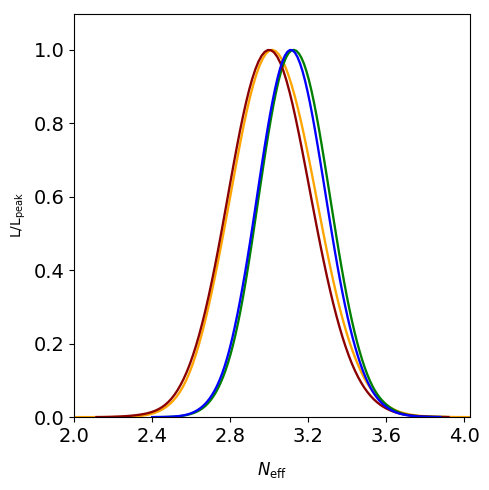} 
\includegraphics[width=2.2in]{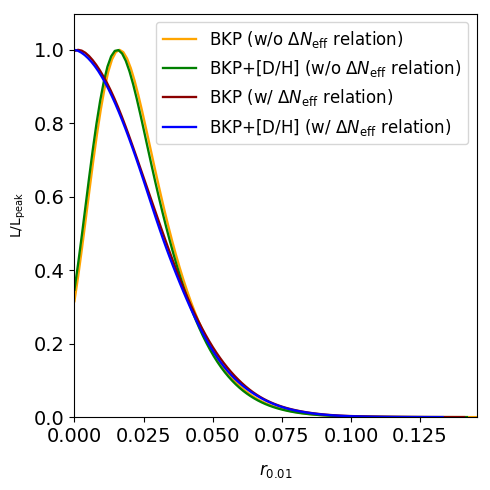}
\includegraphics[width=2.2in]{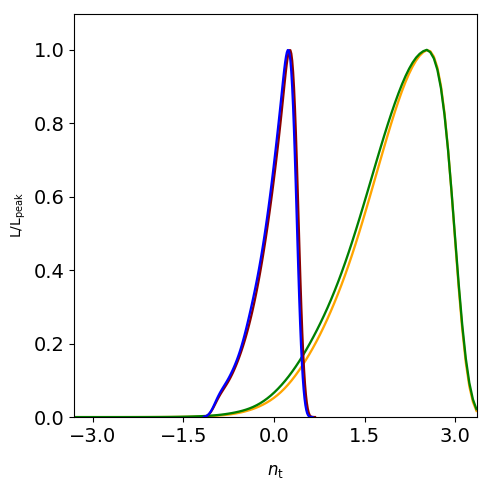}
\includegraphics[width=2.2in]{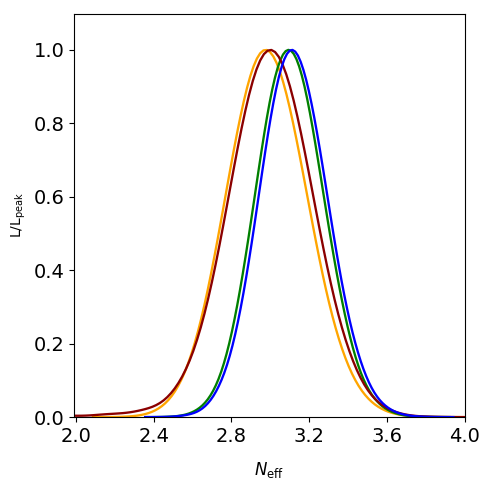}
\caption{The 1-dimensional posterior distributions for the tensor to scalar ratio $r_{0.01}$, the tensor tilt $n_{\rm t}$ and the effective number of neutrino species $N_\mathrm{eff}$ for \textit{Top:} \textit{Planck} dataset and \textit{Bottom:} {\it Planck} + BICEP2 + KECK dataset. The  plots show the distributions for the default case (yellow curve), adding [D/H] data without $\Delta N_\mathrm{eff}$ constraint (green curve), adding $\Delta N_\mathrm{eff}$ constraint (red curve), and adding [D/H] data with $\Delta N_\mathrm{eff}$ constraint (blue curve).} 
\label{fig:1dfig}
\end{figure*}

\begin{table*}[!]
\centering 
\begin{tabular}{c | c | c c | c c} 
\hline\hline 
Parameters & $\Delta N_\mathrm{eff}$ relation & \textit{Planck} & \,\, \textit{Planck}+[D/H] & BKP & BKP+[D/H] \\ [0.5ex] 
\hline
$ \Omega_{\rm b} h^2$ & No & $0.0222 \pm 0.0002$ & $0.0223 \pm 0.0002$ & $0.0222 \pm 0.0002$ & $0.0222 \pm 0.0002$  \\ 
& Yes & $0.0222 \pm 0.0002$ & $0.0222 \pm 0.0002$ & $0.0222 \pm 0.0002$ & $0.0222 \pm 0.0002$ \\ 
\hline
$A_{\rm s}$ & No & $3.09 \pm 0.04$ & $3.10 \pm 0.04$ & $3.09\pm 0.04$ & $3.10 \pm 0.04$ \\ 
& Yes & $3.09 \pm 0.04$ & $3.09 \pm 0.04$ & $3.10\pm 0.04$ & $3.10 \pm  0.04$ \\ 
\hline
$r_{0.01}$ & No & $<0.598$ & $<0.492$ & $<0.070$  & $<0.069$ \\
& Yes & $<0.155$ & $<0.156$ & $<0.073$ & $<0.073$ \\
\hline
$n_{\rm t}$ & No & $0.68 \pm 1.22$ & $0.66 \pm 1.20$ & $2.00\pm 0.72$ & $1.97 \pm 0.74$ \\
& Yes & $-0.64 \pm 0.74$ & $-0.64 \pm 0.74	$ & $0.01 \pm 0.31$ & $-0.01 \pm 0.31$ \\
\hline
$N_\mathrm{eff}$ & No & $3.026 \pm 0.209$ & $3.132 \pm 0.172$ & $2.985 \pm 0.204$ & $3.103 \pm 0.171$ \\
& Yes & $3.005 \pm 0.207$ & $3.118 \pm 0.172$ & $3.000 \pm 0.219$ & $3.122 \pm 0.171$ \\
\hline
$-\log \mathcal{L}$  & No & $5527$ & $5528$ & $5895$ & $5896$ \\ 
& Yes & $5529$ & $5529$ & $5897$ & $5897$ \\ [1ex] 
\hline \hline
\end{tabular}
\caption{Marginalized values of cosmological parameters, effective neutrinos and $\log \mathcal{L}$ values using various combinations of \textit{Planck}, \textit{Planck} + BICEP2 + KECK (BKP) and deuterium datasets, with and without adding the $\Delta N_\mathrm{eff}$ equation. The values quoted for $r_{0.01}$ is the $3\sigma$ upper limit value for the tensor-to-scalar ratio at the pivot scale of $k_{0}=0.01\,\mathrm{Mpc}^{-1}.$} 
\label{table:params} 
\end{table*}

We follow the {\it Planck} 2015 analysis on constraints of inflation \cite{planck2015_inflation}, relax the inflationary consistency relation and use the $(r_{0.01}, n_{\rm t})$ parametrization, where $r_{0.01}$ is the tensor-to-scalar ratio at decorrelation scale $k_{0}=0.01~\mathrm{Mpc^{-1}}$ for the BKP joint constraints. We summarize our joint constraints from CosmoMC runs in Figs.~\ref{fig:bkpver} and \ref{fig:2dfig}, and show the marginalized one-dimensional posteriori distribution in Fig.~\ref{fig:1dfig}. We present the quantitative values in Table~\ref{table:params}. To compare the prediction of the single-field slow-roll inflation model ($r=-8n_{\rm t}$) with the current constraints, we plot this ``consistency relation line'' as black dashed line in the ($r_{0.01},n_{\rm t}$) parameter space in the left and middle panels of Figs.~\ref{fig:bkpver} and \ref{fig:2dfig}.

\subsection{{\it Planck} CMB only}

We first use the CMB data only from \textit{Planck} TT+TE+EE datasets without using the $\Delta N_{\rm eff}$ relation (Eq.~(\ref{eq:delta-Neff})), and show our results in the yellow contours in the left panel of Fig.~\ref{fig:2dfig} and yellow lines in the upper row of Fig.~\ref{fig:1dfig}\footnote{As a matter of consistency check, the yellow contours on the left panel of  Fig.~\ref{fig:2dfig} with \textit{Planck} data only matches fig.~59 in~\cite{planck2015_inflation}.}. One can see that, even without the B-mode polarization data, the \textit{Planck} temperature and E-mode polarization data is already able to put constraints on $r_{0.01}$ as $r_{0.01}<0.60$ at $3\sigma$ C.L. This is because primordial tensor mode can also source the temperature anisotropy before it enters into the horizon, so for $\ell<\ell_{\rm R}$ ($\ell_{R}$ is the multipole (inverse angular size) of the horizon size at recombination), there is a non-negligible contribution to the temperature anisotropy at large angular scales~\cite{Pritchard_2005}. Therefore, cosmic-variance limited measurement of $C^{\rm TT}_{\ell}$, $C^{\rm TE}_{\ell}$ and $C^{\rm EE}_{\ell}$ place constraints on the amplitude of primordial tensor mode.

The red contour in left panel of Fig.~\ref{fig:2dfig} and red line in upper row of Fig.~\ref{fig:1dfig} use the $\Delta N_{\rm eff}$ relation (Eq.~(\ref{eq:delta-Neff})) in the CosmoMC code for {\it Planck} data only case. One can see that with this relation, the parameters of ($r_{0.01},n_{\rm t}$) are tighten up immediately, due to the fact that larger value of resultant $N_{\rm eff}$ can shift the CMB power spectrum (both amplitude and phase) to large extent (see also Fig.~\ref{fig:cls}). As shown in Table~\ref{table:params}, $r_{0.01}$ and $n_{\rm t}$ are tightened to be $r_{0.01}<0.155$ ($3\sigma$ C.L.) and $n_{\rm t}=-0.64 \pm 0.74$. This is a much tighter constraint that the case without $\Delta N_{\rm eff}$ relation (Eq.~(\ref{eq:delta-Neff})).

\subsection{BICEP2 \& KECK Array data 2015 and 2018}

The additional data from the BICEP2/KECK (BK) array which currently has the tightest upper limits on the B-mode power spectrum, would unequivocally improve our constraints on the above tensor parameters. As discussed in Section~\ref{data:cmb}, with the release of both BK 2015 and 2018 data, we evaluated our MCMC runs for both BK releases, and show our results in Figs.~\ref{fig:bkpver} and \ref{fig:2dfig}. One can see from Fig.~\ref{fig:bkpver} that, with BK 2015+{\it Planck} 2015 dataset the $r_{0.01}$ parameter is constrained at $r<0.12$ ($2\sigma$ C.L.), while $n_{\rm t}$ is constrained to be center at zero but have almost equal probability at negative (red) and positive (blue) sides. However, with BK 2018 + {\it Planck} 2015 (BKP) dataset, the constraint on $n_{\rm t}$ clearly prefers a blue-tilted tensor power spectrum at $\sim 2\sigma$ C.L., as shown in the blue contours in Fig.~\ref{fig:bkpver} and yellow line in the middle panel of lower row in Fig.~\ref{fig:1dfig}. This has $2\sigma$ C.L. tension with the consistency relation of single-field slow-roll inflation model. 

In addition to the tilt, the $r$ value is suppressed to be at $r<0.07$ at $2\sigma$ C.L. (Table~\ref{table:params}). This is distinctly evident by the compressed two-dimensional contours for the tensor parameters by adding the BK data to the \textit{Planck} dataset, as seen by comparing the scales of the left and middle panels of Fig.~\ref{fig:2dfig} (yellow contours). 

But with the inclusion of $\Delta N_\mathrm{eff}$ relation (Eq.~(\ref{eq:delta-Neff})), the BKP constraint on ($r_{0.01},n_{\rm t}$) is re-centered at $n_{\rm t}\simeq 0$ with slightly favour over the red-tilt, shown as red contours in the middle panel of Fig.~\ref{fig:2dfig} and red line in the middle panel of Fig.~\ref{fig:1dfig}. The $r$ and $n_{\rm t}$ are constrained to be $r_{0.01}<0.073$ ($3\sigma$ C.L.) and $n_{\rm t}=0.01\pm 0.31$ ($1\sigma$ C.L.) for BKP data set.


\subsection{Deuterium abundance data}
We further include the deuterium abundance measurement~\cite{cooke2018} (Eq.~(\ref{eq:DH})) to tighten up the constraints. Deuterium abundance is sensitive to the baryon density $\Omega_{\rm b}h^{2}$ and $N_{\rm eff}$, which can affect the constraints on $r$ and $n_{\rm t}$ since these parameters are correlated. We first include [D/H] measurement for {\it Planck} data only case in the left panel of Fig.~\ref{fig:2dfig} and upper row of Fig.~\ref{fig:1dfig}. The addition of [D/H] data does not improve the constraints too much.

But the effect of [D/H] data kicks in when we uses the BKP data sets with the inclusion of $\Delta N_{\rm eff}$ relation (Eq.~(\ref{eq:delta-Neff})), as shown in the middle panel of Fig.~\ref{fig:2dfig} and lower row of Fig.~\ref{fig:1dfig}. Comparing the red and blue contours in the middle panel of Fig.~\ref{fig:2dfig}, the upper limit of $r$ is further tightened up with the additional [D/H] data set. The tightest constraints on $r$ and $n_{\rm t}$ become $r_{0.01}<0.073$ ($3\sigma$ C.L.) and $n_{\rm t}=-0.01 \pm 0.31$ ($1\sigma$ C.L.) respectively. We also plot, as the black dashed line, the inflation consistency relation in the middle panel of Fig.~\ref{fig:2dfig}. One can see that the current BKP+[D/H] data with $\Delta N_{\rm eff}$ relation contains this line as its center, indicating that the current data is still consistent with the single-field slow-roll inflation scenario but not excluding other scenarios.

Besides the ($r,n_{\rm t}$) constraints, the addition of the [D/H] dataset can also tighten up $N_{\rm eff}-\Omega_{\rm b} h^2$ joint constraints, as seen in the right panel of Fig.~\ref{fig:2dfig}. [D/H] data provides much tighter constraints to $N_{\rm eff}$ rather than CMB data alone. From Table~\ref{table:params}, we see the positive value of $\Delta N_\mathrm{eff}$ from the CosmoMC runs fall well within the theoretical values as seen in Fig.~\ref{fig:yNeff}. The tightest constraint of $N_{\rm eff}$ is $N_{\rm eff}=3.122 \pm 0.177$ for the current BKP+[D/H] data with $\Delta N_{\rm eff}$ relation, excluding the fourth species of neutrino/relativistic particle at more than $5\sigma$ C.L.

The other cosmological parameters, such as $\Omega_{\rm c}h^{2}$, $100\theta_{\rm MC}$, $A_{\rm s}$, $n_{\rm s}$ and $\tau$ are not affected strongly by including the $\Delta N_{\rm eff}$ relation, so we do not present the results of these parameters here though we release them in the likelihood chain. We refer the interested readers to the references \cite{planck2015_data, bicep, bicep2018}, which are equivalent to our constraints. 


\section{Conclusions and Future Prospects}
\label{conclusions}

In this work, we proposed a new relation between the amplitude ($r$) and the tilt ($n_{\rm t}$) of the primordial tensor power spectrum with the additional effective number of relativistic degree of freedom ($\Delta N_{\rm eff}$). The physics is that the bluer the tilt is, more degree of freedom of stochastic primordial wave background will act {\it like} additional neutrino species, boosting the value of $N_{\rm eff}$ ($N_{\rm eff}=N_{\nu}+\Delta N_{\rm eff}$, where $N_{\nu}=3.046$). This results in more production of the primordial deuterium and enhances the CMB damping tail in the temperature power spectrum. With the combination of the CMB polarization power spectrum, one can place tight and reliable constraints on $r$ and $n_{\rm t}$. 

In this work, we used the {\it Planck} 2015 likelihood chain, combined with BICEP2/KECK (BK) array 2018 release and [D/H] measurement from Damped Lyman-Alpha (DLA) forest system to place constraints on $r$, $n_{\rm t}$ and $N_{\rm eff}$. We first show the results of {\it Planck} only constraints, and then add BK results of 2015 and 2018, and finally add [D/H] measurement, for both the case with and without $\Delta N_{\rm eff}$ relation (Eq.~(\ref{eq:delta-Neff})). One can see that, even without BK result, the inclusion of $\Delta N_{\rm eff}$ relation (Eq.~(\ref{eq:delta-Neff})) significant improves the constraints of ($r,n_{\rm t}$). With additional BK data, the constraints on $r_{0.01}$ and $n_{\rm t}$ are further tightened up. With additional [D/H] data, the tightest constraints can be achieved as $r<0.073$ ($3\sigma$ C.L.) and $n_{\rm t}=-0.01 \pm 0.31$ ($1\sigma$ C.L.). This already places stringent constraints on inflation model, as it still favours the single-field slow-roll inflation model. The tightest constraints of $N_{\rm eff}$ is $N_{\rm eff}=3.122 \pm 0.171$ for BKP2018+[D/H] data, excluding the fourth species of neutrino at high significance.

%

In the future, the combination of direct and indirect measurements from future experiments will put stronger constraints on the stochastic GWs at different frequencies. Future CMB experiments (Simons Observatory~\cite{SO} and COrE~\cite{core}) will improve the measurement of polarisation of CMB and push further down the limit on $r$ and $n_{\rm t}$, or instead of measuring them. With the $\Delta N_{\rm eff}$ relation proposed in this work, measurement of the primordial helium and deuterium abundance will provide stringent constraints on the tensor parameters, indirectly shedding light on to the stochastic GW background. At higher frequency part of the spectrum, direct measurements from GW experiments on a large range of frequencies~\cite{LISA1}, including the ground-based Laser Interferometer Gravitational-Wave Observatory (LIGO)~\cite{ligo} and space-based interferometers like the Laser Interferometer Space Antenna (LISA)~\cite{LISA} will also put constraints on stochastic GW background. The combination of these experiments will result in stringent tests of gravitational radiation and early universe physics.




\section*{Acknowledgements}
We would like to thank Kris Sigurdson for the suggestion at the early stage of this work, Ryan Cooke for discussion on deuterium measurement and primordial abundance, and Antony Lewis, Fabio Finelli and Jussi Valiviita for discussions on {\it Planck} 2015 likelihood and CosmoMC. We would like to thank Antony Lewis for the use of the publicly available numerical codes CosmoMC and CAMB. All CosmoMC runs were done on the UKZN cluster \emph{hippo}. M.A. would like to thank the Claude Leon foundation South Africa for the post-doctoral fellowship granted during the time this work was started, and the South African Radio Astronomy Observatory (SARAO) research fellow grant subsequently. Y.Z.M. acknowledges the support by the National Research Foundation of South Africa with Grant no.105925 and 120378, and the National Science Foundation of China with Grant no.11828301.

%
%
%
%



\end{document}